\documentstyle[prc,aps,twocolumn,epsfig]{revtex}
\newcommand{\ber}{\begin{eqnarray}}
\newcommand{\eer}{\end{eqnarray}}
\begin{document}
\twocolumn[\hsize\textwidth\columnwidth\hsize\csname
@twocolumnfalse\endcsname
\hfill
\title{ A new variable in flow analysis }
\author{Q. H. Zhang}
\address{
Memorial Sloan Kettering Cancer Center, NY, NY 10021}
\author{L. Huo and W. N. Zhang}
\address{
Physics Department, Harbin institute of technology, 15006, P. R.
China }
\date{\today}
\maketitle

\begin{abstract}
We have used a simple spectrum distribution which was derived from a
hydrodynamical equation\cite{Csorgo} to fit the data of the STAR
group. It is found that it can fit the $v_2$ of STAR group very
well. We have found that $v_2$ is sensitive to both the effective
temperature of particles and the expanding velocity. We have
suggested a new variable ${\bf z}$ to be used in the flow analysis.
This new variable will measure the correlation of particles momentum
components. We have also shown that one of the $x$ or $y$ direction
in the reaction plane is the direction which has the largest
variance.
%\vspace{0.5in}
\end{abstract}
]

Recently, the study of elliptic flow has
attracted attention of both
theoreticians and
experimentalists.\cite{flow1,flow2,flow3,flow4,flow5,flow6,flow7,flow8}.
It was argued that an anisotropic distribution
of final state particles with respect to the reaction
plane can be used to reflect the strong re-interaction among
quarks and gluons in the initial state.

The calculation of $v_2$, the anisotropic distribution of final
state particles is always done in the following two ways:(1): we
determine the reaction plane first, then we calculate the average of
$\langle \cos(2(\phi-\phi_R))\rangle$. Here $\phi_R$ is the
azimuthal angle of the reaction plane and $\phi$ is the azimuthal
angle of particles in the Lab frame. (2): using the pairwise
azimuthal correlations\cite{correlation} to calculate the $v_2$. The
second method has an advantage over the first method that no
reaction plane is needed to be determined. But a prior knowledge on
the distribution of $P(\phi)$ is needed to be known for the second
method. The biggest problem for the first method is to determine the
reaction plane. There are several methods are used in data analysis.
In RHIC, the so called second harmonic event plane is widely used in
the data analyzes.

However, it seems to the authors that we still not clear what we
really have measured in the experiment. To discuss this question, we
will use the spectrum distribution in Ref.\cite{Csorgo}. This
spectrum distribution is a
  solution for a hydrodynamical equation and it can be
  expressed as
\begin{equation}
P({\bf p}) \propto \exp(-\frac{(p_x-mv_x)^2}{2mT_x}
                 -\frac{(p_y-mv_y)^2}{2mT_y}).
\end{equation}
Here ${\bf v}=(v_x,v_y)$ is the expanding velocity of particles.
$T_x$ and $T_y$ are effective temperatures of $x$ component and $y$
component of particles momentum\cite{Csorgo}. The above picture can
be understood in the following way: AA collision will form a QGP or
a dense hadron phase at the initial state. Due to the initial
asymmetry collisions, it is expected that the effective temperature
will be different in $x$ and $y$ components. After some time, the
hadron gas are formed.  This hadron gas will expand in velocity
${\bf v}$. This distribution is relative to the reaction plane.

Using this equation, we will calculate $v_2$ using following three
variables. They are ${\bf p}$, ${\bf \tilde{p}}={\bf p}-{\bf
\bar{p}}$, and ${\bf z}=(\frac{\tilde{p}_x}{\langle
(\tilde{p}_x)^2\rangle} ,\frac{\tilde{p}_y}{\langle
(\tilde{p}_y)^2\rangle})$. Here $\langle \tilde{p}_x^2 \rangle$ and
$\langle \tilde{p}_y^2\rangle$ are second moments in $x$ and $y$
direction respectively which are defined as
\begin{eqnarray}
\langle \tilde{p}_x^2 \rangle&=&
\int P({\bf \tilde{p}}) \tilde{p}_x^2 d{\bf \tilde{p}}
\nonumber\\
\langle \tilde{p}_y^2 \rangle&=&
\int P({\bf \tilde{p}}) \tilde{p}_y^2 d{\bf \tilde{p}}.
\end{eqnarray}
Then we will calculate $v_2$ using the following formula:
\begin{equation}
v_2=\frac{\int P({\bf u})
\frac{u_x^2-u_y^2}{u_x^2+u_y^2}
d{\bf u}}{\int P({\bf u}) d{\bf u}}
\end{equation}
Here ${\bf u}$ is one of the three variables in the
above formula. It is easily checked that
\begin{equation}
v_2(z)=0
\end{equation}
for all cases.
\begin{equation}
v_2(\tilde{p})=0
\end{equation}
when $T_x=T_y$; but $v_2(\tilde{p})$ is not zero when $T_x \ne T_y$.
$v_2({\bf p})$ is not zero and its value is shown in Fig.1. It is
interesting to notice that flow may be caused by the expansion
velocity $v$ and the "effective temperature" if we use variables
${\bf p}$ and ${\bf \tilde{p}}$. However, there is no flow to be
observed if we use variable $z$.

\begin{figure}[h]\epsfxsize=8cm
\centerline{\epsfbox{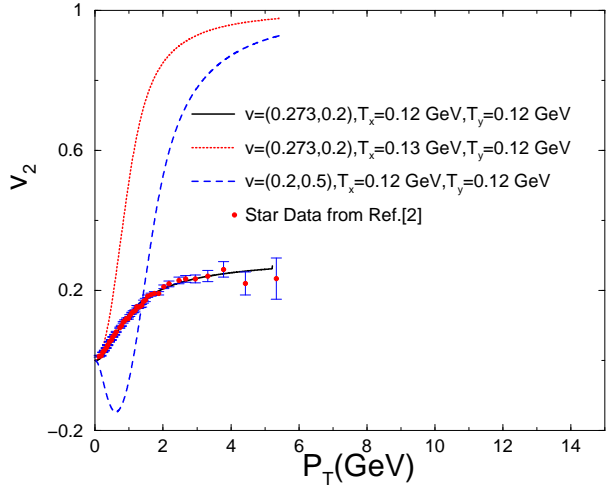}}
%\vskip -2.5cm
\caption{\it
$v_2$ vs. $p_t$.
}
\label{f1}
\end{figure}

From Fig.1, we notice that:(1) the differences between
$T_x$ and $T_y$ can cause strong flow.
(2): The difference between $v_x$ and $v_y$
also cause flow. It seems that the data does not support
the assumption that $T_x$ is different from $T_y$. But we
need to point out that this results is model dependent.
One interesting thing is that this model can fit
the data very well especially in the larger $p_T$ region.

For elliptic flow analysis, we are more interested in the
distribution of particles in momentum space or azimuthal angle
space. We would like to see the relative differences between
$x$-components and $y$-components of particles momentum. The
difference between $T_x$ and $T_y$ might be caused by the initial
flow in QGP phase. But $T_x$ and $T_y$ are measurements which could
tell us the absolutely value of particles momentum in different
direction.  Besides this effective temperatures difference we also
like to find a correlation inside  the distribution of $P({\bf
p_x,p_y})$. From the above simple model, we find that $z$ variable
is a proper variable for this purpose. We would like it can be used
in the data analysis.

To understand the meaning of the flow, we will start from the
transverse momentum sphericity tensor defined by
\begin{equation}
S_{ij}=\sum_{\nu=1}^{M} u_i(\nu)u_j(\nu),
\end{equation}
where $u_i (i=1,2)$ is the $i$-th component of the variable
$u$.  As above, this ${\bf u}$ can be one of three variables
mention above. To be more clearly, we have the following
matrix.
$$
$$
%\begin{matrix}

\[\left[
\begin{array}{cc}
\sum_{\nu=1}^{M} u_1^2(\nu) & \sum_{\nu=1}^{M} u_1(\nu) u_2(\nu)\\
\sum_{\nu=1}^{M} u_1(\nu) u_2(\nu) & \sum_{\nu=1}^{M} u_2^2(\nu)
\end{array}
\right]\]

%\end{matrix}
$$
$$

It is clear that the eigenvalue of this matrix are
\begin{eqnarray}
\lambda_1&=&
\frac{S_{11}+S_{22}+\sqrt{(S_{11}-S_{22})^2+4S_{12}^2 }}{2},
\nonumber\\
\lambda_2&=&
\frac{S_{11}+S_{22}-\sqrt{(S_{11}-S_{22})^2+4S_{12}^2 }}{2}.
\label{e7}
\end{eqnarray}
Its eigenvectors determine the $x$ direction and $y$ direction  of the reaction
plane if we take ${\bf u}={\tilde{p}}$. However
transverse momentum method is widely used in the data analysis to determine
the direction of reaction plane. The reaction plane is determined by
using the following formula:
\begin{equation}
{\bf Q} =\sum_{i=1}^{M} \omega_i {\bf u}.
\label{t1}
\end{equation}
Here $\omega_i =1$ when particles rapidity is big than zero and
$\omega_i =-1$ when particles rapidity is less than zero. The basic
idea behind this method is that the total transverse momentum
of the system in the $y$ direction of the
reaction plane is zero and the symmetry of the collision system
in the direction of $\theta$ to $\theta +\pi$. Here $\theta$ is
the polar angle. Can this two methods give the same reaction plane?

We will give a positive answer to this question under the condition
that the multiplicity of event is infinity in this
paragraph. When we determine the two eigenvectors, then
$S_{12}=0$ in this new frames.
 If we use Eq.(\ref{t1}) to determine
the reaction plane first, then we can also find that $S_{12}$ will
be zero in the frame  due to the symmetry of $u_2 \leftrightarrow
-u_2$.  We will show that there is only one vector in the $x-y$
planes which lead the value of $S_{12}$ is zero. Suppose that a new
axis which has angle relative to the reaction plane is $\theta$. It
is easily checked that the new quantity in this new frame is
\begin{eqnarray}
u_x'(\nu)&=&u_x^{R}(\nu)\cos(\theta) + u_y^{R}(\nu) \sin(\theta)
\nonumber\\
u_y'(\nu)&=&-u_x^{R}(\nu)\sin(\theta) + u_y^{R}(\nu) \cos(\theta)
\end{eqnarray}
Here the superscript $R$ refers that those quantities are
in the reaction plane frame. ${\bf u'(\nu)}$ is the variable
in the new frame.  Then
\begin{equation}
S_{12}'=\sum_{\nu=1}^{M} u_x'(\nu)u_y'(\nu)=
\frac{ [S_{22}^{R}-S_{11}^{R}]\sin(2\theta)}{2}.
\end{equation}
It is clear that $S_{12}'$ is zero only when $\theta=0$ or $\pi$
or $S_{11}^{R}=S_{22}^{R}$. For the first case, it means that
$S_{12}'=0$ when it is in
the reaction plane frame. When $S_{11}^R=S_{22}^R$,
the shape of $u^{T}Su$ is circle in momentum space.
The eigenvector for this matrix is arbitrary.
Therefore, we have shown that the
transverse momentum method and sphericity tensor give the same reaction plane
when the multiplicity is huge.

Then $\alpha$ which is used to measure the
 asymmetry distribution
of particles in the momentum
can be calculated in the following ways.
\begin{equation}
\alpha=\frac{\lambda_1-\lambda_2}{\lambda_1+\lambda_2}
=\frac{(\sum_{\nu=1}^{M} u_x^R(\nu))^2-
(\sum_{\nu=1}^{M} u_y^R)^2}
{(\sum_{\nu=1}^{M} u_x^R(\nu))^2+(\sum_{\nu=1}^{M} u_y^R(\nu))^2}
\end{equation}
Using Eq(\ref{e7}), we have
\begin{equation}
\alpha =\frac{\sqrt{(S_{11}-S_{22})^2 +4S_{12}^2}}{S_{11}+S_{22}}.
\label{e12}
\end{equation}
Taking ${\bf u=p_t}$,
$\alpha$ can be expressed as
\begin{equation}
\alpha = \frac{1}{M}\sum_{\nu=1}^{M} \cos(2\phi_{\nu}^R) \frac{p_t(\nu)^2}{\langle p_t^2\rangle}
\end{equation}
Here $\langle p_t^2\rangle=\frac{1}{M}\sum_{\nu=1}^{M} {\bf p_t^2}$ are
the transverse momentum square average for
the event.  $M$ is the multiplicity for the event. Then we have
\begin{equation}
\langle \alpha \rangle =
\langle \cos(2\phi_{\nu}^R)
\frac{p_t(\nu)^2}{\langle p_t^2\rangle} \rangle.
\end{equation}
This average is taken over all events.
If the transverse momentum is fixed, then we have
\begin{equation}
\alpha(p_t)=v_2(p_t).
\end{equation}
Therefore, we
can calculate $v_2(p_t)$ using Eq.(\ref{e12}) directly.
If we choose ${\bf u}={\bf z}=(z_1,z_2)=
(\frac{\tilde{p_x}}{\sum \tilde{p_x}^2},
\frac{\tilde{p_y}}{\sum \tilde{p_y}^2})$, then we find that
the matrix for $z$ becomes to
$$
$$
%\begin{matrix}
\[\left[
\begin{array}{cc}
 1 & \sum_{\nu=1}^{M} z_1(\nu) z_2(\nu)\\
\sum_{\nu=1}^{M} z_1(\nu) z_2(\nu) & 1
\end{array}
\right]\]
%\end{matrix}
$$
$$
Then the $\alpha$ for this new variable will be
\begin{equation}
\alpha ({\bf z})=|S_{12}(z)| =\frac{|S_{12}|}
{\sqrt{S_{11}S_{22}}}.
\end{equation}
When particles rapidity window is huge and number of particle is
infinity for each event, we can write
\begin{eqnarray}
S_{11}&=&\int P({\bf \tilde{p}}) {\tilde{p}_x} {\tilde{p}_x} d{\tilde{p}_x} d{\tilde{p}_y}
\nonumber\\
S_{22}&=&\int P({\bf \tilde{p}}) {\tilde{p}_y} {\tilde{p}_y} d{\tilde{p}_x} d{\tilde{p}_y}
\nonumber\\
S_{12}&=&\int P({\bf \tilde{p}}) {\tilde{p}_x} {\tilde{p}_y} d{\tilde{p}_x} d{\tilde{p}_y}
\end{eqnarray}
Therefore
%{\footnote{ We need to point out that all those quantities
%such as ${\tilde{p}}$ and $P({\tilde{p}})$ are in the lab frame.
%They are also quantities for a event. If we have the symmetry of
%${\tilde{p_y}}\leftrightarrow -{\tilde{p_y}}$, then only half the
%phase space will be used in the calculation.} }
\begin{equation}
\alpha(z)=
\frac{
|\int P({\bf \tilde{p}}) {\tilde{p}_x} {\tilde{p}_y} d{\tilde{p}_x} d{\tilde{p}_y}|}
{\sqrt{\int P({\bf \tilde{p}}) {\tilde{p}_x} {\tilde{p}_x} d{\tilde{p}_x} d{\tilde{p}_y}
\int P({\bf \tilde{p}}) {\tilde{p}_y} {\tilde{p}_y} d{\tilde{p}_x} d{\tilde{p}_y}}}.
\end{equation}
Thus $\alpha(z)$ actually measures the correlation between the two component
of particles. If the particles are emitted randomly, $\alpha(z)=0$.
On the other hand, if particles are emitted always in the a
particular direction, say $p_x=p_y$, then $\alpha(z)=1$.
It is easily checked that
\begin{equation}
\alpha(\tilde{p})=\sqrt{1+4\frac{t}{(1+t)^2} (\alpha(z)^2-1)}.
\end{equation}
Here $t=\frac{S_{11}}{S_{22}}$. It is interesting to notice that
when $t=1$, then $\alpha({\tilde{p}})=\alpha(z)$.
It is easily seen that
$\alpha({\tilde{p}})\ge \alpha(z)$. If $t=0$, then
$\alpha({\tilde{p_t}})=1$ which corresponds
to the case particles are emitted only in the $y$ direction. The advantage
of $\alpha(z)$ over $\alpha({\tilde{p_t}})$ is clear:
it depends on a dimensionless variable; it measures the correlations
between the components of particles momentum.

If we take ${\bf u=p_t}$ and
${\bf y}=
(\frac{p_x}{\langle p_x^2\rangle },\frac{p_y}{\langle p_y^2\rangle})$, then we
will have similar expressions as above.  The only difference is that we
need to use varible ${\bf p_t}$ and ${\bf y}$ to take the place of
variables ${\tilde{p_t}}$ and ${\bf z}$ in above expressions.

If $\alpha(z)$ is very small, we have
\begin{equation}
\alpha(\tilde{p})\sim \frac{|1-t|}{1+t}=\frac{|S_{11}-S_{22}|}{S_{11}+S_{22}}.
\end{equation}
Thus $\alpha({\tilde{p}})$ measure the ratio between the
difference of the variance in $x$ and $y$
directions and the sum of the variance in $x$ and $y$ directions.
The detail information of $\alpha(z)$ will not be observed.

One of interesting things about the matrix of variable $z$ is that its
eigenvector is always along the direction
$(\frac{1}{\sqrt{2}},\frac{1}{\sqrt{2}})$
or $(\frac{1}{\sqrt{2}},\frac{-1}{\sqrt{2}})$.
Due to the fact that the finite multiplicity
in the collisions, the "estimated reaction plane" will
 fluctuate randomly.  Finally, we need to point out that
this estimated reaction plane is quite different from the reaction
plane estimated using variable ${\bf \tilde{p}}$
 since they are
in different variable spaces.

In the experiment, experimentalists normally used
${\bf u}=\frac{\bf{p_t}}{|{\bf p_t}|}=(\cos(\phi),\sin(\phi))$
in the analysis. The matrix for $x$ is
$$
$$
%\begin{matrix}
\[\left[
\begin{array}{cc}
\sum_{\nu=1}^{M}\cos^2(\phi_\nu) & \sum_{\nu=1}^{M} \cos(\phi_\nu) \sin(\phi_\nu)\\
\sum_{\nu=1}^{M} \cos(\phi_\nu) \sin(\phi_\nu) & \sum_{\nu=1}^{M} \sin^2(\phi_\nu)
\end{array}
\right]\]
%\end{matrix}
$$
$$
When the system transform to  eigenvectors frame, we have
\begin{equation}
\sum_{\nu=1}^{M}\sin(\phi_\nu-\phi_R) \cos(\phi_nu-\phi_R)
=\frac{1}{2}\sum_{\nu=1}^{M} \sin2(\phi_{\nu}-\phi_R)=0.
\end{equation}
This is the way to calculate the reaction plane
azimuthal angle ($\phi_R$) in RHIC data analyses.

We can construct other varibles to dertermine the high-order
harmonics reaction plane. For example, when $n=4$, we can take
${\bf u}=( \frac{2p_xp_y}{|{\bf p_t}|^2},
\frac{p_x^2-p_y^2}{|{\bf p_t}|^2})=(\sin(2\phi),\cos(2\phi))$.
If we calculate  the matrix in
 the corresponding eigenvector frame, we
have
\begin{eqnarray}
&&\sum_{\nu=1}^{M}\sin2(\phi_{\nu}-\phi_R)\cos(2(\phi_{\nu}-\phi_R))
\nonumber\\
&=&\frac{1}{2}\sum_{\nu=1}^{M} \sin4(\phi_{\nu}-\phi_R)=0.
\end{eqnarray}
This is the way to calculate the fourth harmonic reaction plane
as metioned in the Ref.\cite{flow9}. In general this estimated
reaction plane should be different from the estimated
second harmonic reaction plane since they are estimated for different
variables. The corresponding $\alpha$ for this matrix is $v_4$.

We have shown in the above that when the multiplicity is huge, transverse
momentum and spherecity matrix will give the same "estimated" reaction
plane for variable ${\bf u}$ as long as they have symmetry
$u_y\leftrightarrow -u_y$. However, we know that this estimated reaction plane
is quite different for different
variable ${\bf u}$.  What is the physical meaning of the
direction estimated by sphericity matrix and transverse momentum? We will
show in the next paragraph that this eigenvectors directions actually
gives the direction where the variance is the maximum.

Suppose that ${\bf u}$ is a vector ($2\times 1$ matrix) in the lab frame and we
will try
to find a new frame such that the new variable ${\bf u'}$ has the
largest variance along a axis in the frame. Suppose $a^T$ is a $2\times 2$ matrix,
then the new variable $u'$will be
\begin{equation}
u_{i}'={\bf a_i^T} {\bf u}, ~~~~~~~i=1,2.
\end{equation}
Here ${\bf a_i^T}=(a_{i1},a_{i2})$.
Then its variance is
\begin{equation}
\sum_{\nu=1}^{M} (u_{i}'(\nu))u_i'^T(\nu)=
{\bf a_i^T u u^T a_i}={\bf a_i^T} S {\bf a_i}, ~~~~~~~i=1,2.
\end{equation}
if we take a constraint that ${\bf a_i^Ta_i}=1$. Then we can
construct the
following Lagrange multipliers
\begin{equation}
L({\bf a_i})={\bf a_i^T} S {\bf a_i} -\lambda ({\bf a_i^Ta_i} =1).
\end{equation}
Taking derivation with ${\bf a_i}$, we have
\begin{equation}
\frac{\partial L }{\partial {\bf a_i}}= 2 S {\bf a_i} - 2\lambda {\bf a_i}
\end{equation}
Setting this value to zero, we have
\begin{equation}
S {\bf a_i} =\lambda {\bf a_i},
\end{equation}
and
\begin{equation}
\sum_{\nu=1}^{M} (u_{i}'(\nu))^{T}u_i'(\nu)=\lambda_i; i=1,2
\end{equation}
This tell us that when we choose one of the eigenvector
as $a_1$, then its maximum variate will be the eigenvalue. We
will choose the largest eigenvalue. It is also easily
to prove that if we choose $a_2$ as another eigenvalue. Then
the $S_{12}$ will be zero for variable $u'$. Therefore we have shown
here that the one of the eigenvectors actually gives us the direction
where the variance is the largest. For another eigenvector
which give us another direction which has a property that
$S_{12}$ will be zero (in other words, its component will be
uncorrelated with the first component). The largest
variance for our case is the largest eigenvalue.

Conclusions: It has been shown that the $v_2$ measured in the
data reflects the information on the expanding velocity,  effective
temeparature in $x$ and $y$ components and correlation
between different components of the particles momentum.
However due to the fact that the correlation between
different momentum of particles are small. Therefore we
redefine a new variable $z$ which can be used in the
data analysis. We belived that $v_2(z)$
is small but can show us the information on the
correlation between different components of particles.
We have also shown that estimated reaction plane will be
different if we choose different variables in the data
analysis. The two directions in the reaction plane can be
undersood as the biggest variance directions of the
data.
One of the interesting thing is that, the model
can fit the data quite well. This model
suggests that the effective temparature for different
directions of momentums are almost the same.

The authors thank R. C. Hwa, S. Padula for for helpful discussions.
 This work was  supported by the National Natural Science
 Foundation
 of China under Contract No.10275015" . QHZ thank C. Gale for his
 help during the preparation of the paper.

%\section*{Acknowledgments}

%This work was partly supported by the NSERC of Canada and
%FCAR of the Quebec Government.

%\section*{References}
{}

\begin{thebibliography}{10}

\bibitem{flow1}
S. S. Adler et al., Phys. Rev. Lett. 91 , 182301 (2003).
\bibitem{flow2}
J. Adms et. al., Phys. Rev. Lett. 92, 052302 (2004);
J. Adms et. al., Phys. Rev. Lett. 92, 062301 (2004);
C. Adler et. al., Phys. Rev. Lett. 90, 032301 (2003);
C. Adler et. al., Phys. Rev. C 66, 034904 (2002);
C. Adler et. al., Phys. Rev. Lett. 89, 132301 (2002);
C. Adler et. al., Phys. Rev. Lett. 89, 182301 (2001);
C. Adler et. al., Phys. Rev. Lett. 86, 402 (2001).
\bibitem{flow3}
B.B. Back et al., nucl-ex/0406021.
B.B. Back et al., Phys. Rev. Lett 89, 222301 (2002).

\bibitem{flow4}
J. Barrette et al., Phys. Rev. Lett. 70, 2996 (1993),
Phys. Rev. C 56, 3254 (1997),
Phys. Rev. C 55, 1420 (1997).

\bibitem{flow5}
P. Chung et al., Phys. Rev. C 66, 021901 (2002);
\bibitem{flow6}
M. Gyulassy and L. McLerran, nucl-th/0405013.

\bibitem{flow7}
P. F. Kolb and U. Heinz, nucl-th/0305084;
U. Heinz and S. M. H. Wong,
  Phys. Rev. C 66, 014907 (2002).
\bibitem{flow8}
J. Y. Ollitrault, Phys. Rev. D 48 ,1132 (1993),
                 Phys. Rev. D 46, 229 (1992).

\bibitem{flow9}
A. M. Poskanzer and S. A. Voloshin, Phys. Rev. C 58, 1671 (1998).
\bibitem{flow10}
Z. W. Lin and D. Molnar, Phys. Rev. C68, 044901 (2003);
Z. W. Lin and C. M. Ko, Phys. Rev. Lett. 89, 202302 (2002);
D. Monar and S. Voloshin, Phys. Rev. Lett. 91, 092301 (2003).

\bibitem{correlation}
S. Wang et al., Phys. Rev. C 44, 1091 (1991);
R. Lacey et al., Phys. Rev. Lett. 70, 1224 (1993).

\bibitem{Csorgo}
M. Csanad, T. Csorgo and B. Lorstad, nucl-th/0310040;
T. Csorgo, F. Grassi, Y. Hama, T. Kodama, Phys. Lett. B565,107 (2003);
%P. Csizmadia, T. Csorgo and B. Lukacs, Phys. Lett. B443, 21 (1998);
T. Csorgo, S.V. Akkelin, Y. Hama, B. Lukacs and Y. M. Sinyukov,
Phys. Rev. C67, 034904 (2003).

\end{thebibliography}
\end{document}